\documentclass[12pt]{article}
\pdfoutput=1
\usepackage{amsfonts,amsthm,amsmath,amssymb}
\usepackage{hyperref}
\usepackage{cite}
\usepackage[paper=letterpaper,margin=1in]{geometry}
\usepackage{graphicx}
\usepackage{units}

\def\a'{\alpha'}

\renewcommand\bar{\overline}

\newcommand{\be}{\begin{equation}}
\newcommand{\ee}{\end{equation}}
\newcommand{\bea}{\begin{eqnarray}}
\newcommand{\eea}{\end{eqnarray}}
\newcommand{\ba}{\begin{array}}
\newcommand{\ea}{\end{array}}
\newcommand{\ben}{\begin{enumerate}}
\newcommand{\een}{\end{enumerate}}
\newcommand{\bi}{\begin{itemize}}
\newcommand{\ei}{\end{itemize}}
\newcommand{\bc}{\begin{center}}
\newcommand{\ec}{\end{center}}
\newcommand{\bfig}{\begin{figure}}
\newcommand{\efig}{\end{figure}}

\numberwithin{equation}{section}

\begin{document}
\thispagestyle{empty}
\renewcommand{\thefootnote}{\fnsymbol{footnote}}

\begin{flushright}
%{\footnotesize WITS-CTP-149}
\end{flushright}
\vspace{1 cm}

\begin{center}

\bf{{\LARGE Holographic $c$-Function}\\

\vspace{1cm}}

{S. Shajidul Haque\,\footnote[1]{{\tt{sheikh.haque@uct.ac.za}}}}\\

\vspace{0.5 cm}{{\it$^{b}$Laboratory for Quantum Gravity \& Strings\\
Department of Mathematics \& Applied Mathematics\\
University of Cape Town,
South Africa }} \\

\vspace{2cm}
{\bf Abstract}
\end{center}
\begin{quotation}
\noindent We propose a simple and generic holographic $c$-function that is defined purely from geometry by using the non affine expansion for null congruences. We examined the proposal for BPS black solutions in $\mathcal{N}=2$ gauged supergravity that interpolate between two different dimensional AdS spacetimes and also for domain wall solutions. Moreover, we commented on the relation of this geometric proposal with the one from the holographic entanglement entropy. 
\end{quotation}

\setcounter{page}{0}
\setcounter{tocdepth}{2}
\newpage
% \tableofcontents
%%%%%%%%%%%%%%%%%%%%%%%%%%%%%%
%%%%%%%%%%%%%%%%%%%%%%%%%%%%%%
%%%%%%%%%%%%%%%%%%%%%%%%%%%%%%
\section{Introduction}
In two dimensions, conformal fixed points that are connected by a renormalization group flow, there exists a positive real function known as the $c$-function of the coupling constants and energy scale. The value of the $c$-function does not increase along the RG flow. At the fixed points of the flow this $c$-function is stationary and assumes values equal to the central charges of the corresponding conformal field theories~\cite{Zamolodchikov:1986gt}. Since the UV boundary value of the $c$-function is larger than that of IR, the RG evolution of this function is a gradient flow in which the number of degrees of freedom decreases.

The 2d central charge is related to the conformal anomaly, $\langle T^\mu_\mu \rangle = \frac{c}{24 \pi} R $, where $R$ is the Ricci scalar. In four-dimensional an analogue of the $c$ can be obtained from the trace anomaly %vacuum expectation value of the trace of the stress-energy tensor
\be
\langle T^\mu_\mu \rangle \sim  -a\, E_4 + c\, W_{\mu\nu\rho\sigma} W^{\mu\nu\rho\sigma} ~, \label{eulerweyl}
\ee
where $E_4$, the Euler density, is quadratic in the Riemann tensor and $W_{\mu\nu\rho\sigma}$ is the Weyl tensor. In all known examples, $a_\mathrm{UV} \ge a_\mathrm{IR}$, but the same is not true of the coefficient $c$. % in~[\ref{eulerweyl}]. 
Therefore, `$a$' is a possible candidate for the $c$-function in four dimensions \cite{Cardy:1988cwa, Osborn:1989td}. And indeed a few years ago Komargodski and Schwimmer \cite{Komargodski:2011vj,Komargodski:2011xv} established the $a$-theorem in four dimensions.

The AdS/CFT duality motivates us to map the field theory $c$-theorem to a statement about gravity~\cite{Girardello:1998pd, Freedman:1999gp, Myers:2010xs, Paulos:2011zu}. For the gravity theory the $c$-function should imply a geometric quantity that interpolates between two vacuum states in UV and IR. In this paper, we propose a very simple and a generic holographic c-function that is defined entirely from geometry. A similar approach was implemented before in \cite{Sahakian:1999bd}. They also used similar geometric quantity `expansion for null geodesic congruences, $\theta$' as ours. However, their $\theta$ is limited to an affine choice of the null tangent vector, while in this current note we generalize the prescription for non-affine null tangent vectors. 

The motivation for generalizing the $c$-function for non-affine null tangent vector stems from the fact that the field theory $c$-function is not unique and the relation between $c$-function and the $\beta$-function depends on the renormalization scheme\footnote {The fact that there exists a $c$-function is, however, scheme independent \cite{PhysRevD.40.1964}.} \cite{PhysRevD.40.1964, Nojiri:2000sp}. For instance, Zamolodchikov's original proof \cite{Zamolodchikov:1986gt} is based on dimensional regularization with minimal subtraction method, where the gradient of the $c$ function is proportional to the $\beta$-function. Since the $\beta$-function is scheme dependent, it is not surprising that the $c$-function is not unique. Moreover, scheme dependent holographic RG and its implications on holographic $c$-function are discussed in \cite{Erdmenger:2001ja, Anselmi:2000fu}. Now, the holographic $c$-function defined from the affine $\theta$ (as in \cite{Sahakian:1999bd}) is quite unique, since the affine null vector is unique \footnote{Up to some global scaling and translation.} and it fixes the expansion $\theta$ and thus the $c$-function. %It would depend on the energy scale. 
On the other hand our non-affinely defined $c$-function has more freedom as we can choose a family of null vectors and get slightly different expansion parameters. In this sense our proposal
realizes this non-unique nature of the $c$-function unlike the proposal made in \cite{Sahakian:1999bd}.
 %captures the non-unique nature of the $c$-function unlike the proposal made in \cite{Sahakian:1999bd}.

The $c$-function encodes information loss and consequently irreversibility of RG flows in field theories. Therefore, in holographic setups when the aim is to define a dual $c$-function in the gravitational picture, one must look for a quantity that naturally displays geometrical irreversibility. A natural choice for this encoding is the expansion $\theta$ that captures the convergence of null geodesics. The convergence is captured by the derivative of $\theta$ with respect to an affine parameter and it is extremized on a null surface such as a black hole horizon, which is the locus of points where information loss is maximized with respect to the asymptotic observer. This corresponds to an IR fixed point in the field theory side. Motivated by this, we propose and examine a c-function defined in terms of the convergence of null geodesics. 

%
%%%
There are also other holographic proposals for the $c$-function \cite{Cremades:2006ke, Freedman:1999gp, Girardello:1998pd, Anselmi:2000fu, Erdmenger:2001ja, Goldstein:2005hq, me, Myers:2010tj}, some of which are not generic. The proposal \cite{Cremades:2006ke} is not a true holographic dual of the field theory $c$-function.
For spherically symmetric static configurations in four dimensional two derivative gravity, a $c$-function was proposed in~\cite{Goldstein:2005hq}. Their proposal is simply the area of radial slices and consequently insufficient, as it is not well defined at infinity. Another recent proposal \cite{me} establish that for AdS spaces with supersymmetric single centered black holes, the $c$-theorem is a consequence of the attractor mechanism~\cite{Ferrara:1995ih}. Nevertheless, our geometric proposal is a way to generalize the spherically symmetric metric information to define monotonic gradients without using the equations of motion explicitly.

The organization of this paper is as follows. In section two we will characterize the properties of the $c$-function and present our proposal. Then we will justify the proposal based on these properties. In section three we will show some explicit example in gauged supergravity models, but the analysis is otherwise completely general. Finally, in section four we will compare our $c$-function with the proposal based on holographic entanglement entropy. 
%%%%%%%%%%%%%%%%%%%%%%%%%%%%%%
%%%%%%%%%%%%%%%%%%%%%%%%%%%%%%
%%%%%%%%%%%%%%%%%%%%%%%%%%%%%%
\section{Holographic $c$-Function}
For investigating the holographic c-function it is useful to make a list of properties that we want the c-function to satisfy. Some of these properties are used to define the $c$-function. However, there are properties we can identify that are motivated from structure of the boundary value. A similar approach was taken in \cite{Sahakian:1999bd}. The properties are:
\begin{itemize}
\item The c-function should be proportional to the inverse of the gravitational coupling $G^{-(d+1)}$. 
\item It will be dimensionless, therefore, proportional to $L^{d-1}$. 
\item Invariant under boundary diffeomorphism.
\item In AdS spaces it should be constant proportional to $L^{d-1}/G_{d+1}$.
\item Well defined/finite on both sides of the RG flow.
\item Stationary at the end points of RG flow. 
\item Monotonic along the RG flow.
\item It is a positive real function.
\end{itemize}
Next we will describe our new proposal for the holographic c-function and then explain the justifications for the choice.
\\ \\
{\bf $c$-Function and its Justification:}
Let's consider a $d+1$ dimensional spacetime with metric $g_{ab}$ foliated by a choice of $d-1$ dimensional constant time surfaces. We can immediately find the expansion for null geodesic congruences. Using this as the basic geometric ingredient we propose the following c-function:
\be \label{proposal}
c= (-1)^{d-1} \frac{ c_1}{\langle \theta ^{d-1} \rangle }+c_2,
\ee
where $c_1$ and $c_2$ are constants that can be fixed by the central charges at the end points of RG flow. The exact form of these parameters depends on explicit examples. These parameters essentially take care of the presence of appropriate Newton's constant. The quantity $\theta$ is the expansion parameter.  The notation $\langle ... \rangle$ implies an average, taken over co-dimension two surfaces with fixed time and is defined as follows:
\be
\langle \theta ^{d-1} \rangle= \frac{ \int \sqrt {h} \ \theta^{d-1}   } { \int \sqrt {h}  }.
\ee
Here $h$ is the determinant of the metric for the $d-1$ surfaces. The power $d-1$ is also important since this gives us the correct power of AdS length scale. For flow between different dimensional CFTs the appropriate power of the AdS length scale on the infrared boundary will be secured by the constants $c_1$ and $c_2$. Note that the averaging of $\theta^{d-1}$ is crucial for a proposal of the $c$-function. This is because in addition to ensuring diffeomorphism invariance on the boundary it also makes sure we do not get any extra scalar term from the metric.
The appearance of such extra terms is potentially dangerous for the boundary values of the $c$-function. The regularity and stationary value of the c-function is a consequence of the fact that the expansion is a function of ratio of the metric warp factor and harmonic function and their derivatives. On the AdS these scalars become proportional to the radial coordinate which makes the boundary values regular. These issues will be elaborated in the next section when we discuss some explicit examples. The prefactor of the first term on the right hand side of equation \ref{proposal} makes sure that the $c$-function remains positive. 
%%%%%%%
%%%%%%%

Now to prove monotonicity we will use the Raychaudhuri equation. We are interested in the behavior of the expansion $\theta$ of inwardly directed, future-oriented null rays $k^\alpha$. For an affinely-parametrized null tangent vector $k^\alpha \nabla_\alpha k^\beta = 0$, 
the expansion $\theta = \nabla_\alpha k^\alpha$ satisfies the null Raychaudhuri equation:
\be
\frac{d\theta}{d\lambda} = -\frac{1}{2} \theta^2 - |\sigma|^2 - T_{\alpha \beta} k^\alpha k^\beta\, ,
\label{Raychaud_Affine}
\ee
where $\lambda$ is an affine parameter, $\sigma$ is the shear tensor, and we used Einstein's equations.
If we assume that the NEC is satisfied so that $T_{\alpha \beta} k^\alpha k^\beta \geq 0$, the (affine) Raychaudhuri
equation \ref{Raychaud_Affine} implies that the expansion must be non-increasing,
\be
\frac{d\theta}{d\lambda} \leq 0 \, .
\label{RayNoGo}
\ee
%%%
It is sometimes more convenient to parametrize the inwardly directed null rays with non-affine tangent vectors \cite{Blau}, 
$\tilde k^\alpha \nabla_\alpha \tilde k^\beta = \kappa \tilde k^\beta$, where $\kappa$ is the non-affine parameter.
In this case, the divergence 
${\tilde\theta} = \nabla_\alpha \tilde k^\alpha - \kappa$ satisfies a modified form of Raychaudhuri's equation:
\be
\frac{d\tilde \theta}{d\lambda} = \kappa \tilde \theta -\frac{1}{2} \tilde \theta^2 - |\sigma|^2 - T_{\alpha \beta} \tilde k^\alpha \tilde k^\beta\,.
\label{Raychaud_NonAffine} 
\ee
%%%
%%%
For negative $\tilde \theta$ \footnote{This comes from the condition of convergence \cite{Bousso:1999xy}. This was also used in \cite{Sahakian:1999bd}.} we immediately get 
$\frac{d\tilde \theta}{ d\lambda} \leq 0$. We can rewrite the above expression as
\be
\frac{d \tilde\theta}{ d\lambda} = \frac{d \tilde \theta}{ dr} \left (\frac{dr}{d\lambda} \right) =  k^r \frac{d \tilde \theta}{ dr} \leq 0 \, .
\ee
This expression will ensure the  monotonicity of the c-function. In this paper we will be using non-affine null vector, therefore, $\tilde \theta$ in our definition for the $c$-function.
%%%%%%%
%%%%%%%
\section{Examples}
In this section we will explore our proposal for some concrete examples. Note that our definition is based on the averaged expansion parameter which is absolutely a geometric quantity. Therefore, this proposal can be easily applicable to any theory.
\subsection*{$\mathcal{N}=2$ Gauged Supergravity}
We are interested in extremal solutions of $\mathcal{N}=2$ gauged supergravity in $AdS_{d+1}$. These models have been extensively studied over the last few years \cite{Cacciatori:2009iz, Dall'Agata:2010gj, Barisch:2011ui, Barisch-Dick:2013xga, Almuhairi:2011ws, Cucu:2003bm, Benini:2013cda}. Recently, a complete construction of static 4d BPS solutions for all symmetric models is done by \cite{Katmadas:2014faa, Halmagyi:2014qza}. 
The near horizon geometry of the extremal background is of the form $AdS_3 \times \Sigma$ or $AdS_2 \times \Sigma$, where $\Sigma$ can be spherical, hyperbolic or planar depending on the structure of the black hole (brane) horizon. The $d+1$ dimensional static spherically symmetric metric looks like 
\be \label{BH}
ds^2= -a(r)^2 dt^2 +a(r)^{-2} dr^2 +b(r)^2 \sum_i ^{d-2} dx_i^2 +w(r)^2 dz^2.
\ee
Here $r$ is the radial direction corresponding to the RG flow and $\{t, x_i, z\}$ are the boundary coordinates. The harmonic function $a(r)$ and the warp factors $b(r)$ and $w(r)$ that appear in \ref{BH} describes a black hole spacetime in the infrared and give an asymptotically $AdS_{d+1} $ geometry for large $r$. In the following, we will study our proposed c-function for $5d$ and $4d$ solutions respectively.
%%%%
\subsubsection*{$AdS_3$ Near Horizon Geometry}
Let's consider the case where the boundary is $AdS_5$ and the near horizon is $AdS_3 \times \Sigma$. To get such near horizon geometry we will set $w(r)=a(r)$. We will use this example to investigate both our proposal and that of \cite{Sahakian:1999bd} in a generic format. 
We consider a radial inwardly directed null ray. These rays can be described by the null tangent vector
\be
k^\alpha =\frac{dx^{\alpha}  } {d\lambda   }= f(r) \left( a^{-1}, -a,0,0,0\right) \, ,
\ee
where $f(r)$ is a conformal factor which we will use to switch between affine and non-affine vectors. With this choice of the null tangent vector we can immediately find  
\be
%\theta=
\nabla_{\mu} k^{\mu} =-\left(2 \frac{f a b' }{b} + a' f + a f'\right )\, ,
\ee
where the derivative is taken with respect to $r$. The null vector we considered is not affine. For a quick check we need to evaluate the following quantity as discussed in section 2:
\be
k^{\mu} \nabla_{\mu} k^{\nu} = -\left( a'f +f a'  \right )k^{\nu} \, .
\ee
This implies that the quantity $-(a' f+af')$ is essentially the non-affine parameter $\kappa$. 
Therefore, the non-affine expansion can be written as
\be 
\tilde \theta =-2 \frac{a f}{b} b'  \, .
\ee
We can summarize the conditions for affine and non-affine as follows:
\bea \label{conditions}
\text{Affine}: &&\ a f'+ a'f =0 \ \implies  af = \text{constant} \cr
\text{Non-Affine}: &&\ a f'+ a'f \neq 0 \ \implies  af \neq \text{constant} \,.
\eea
Note that for the non-affine case, we have some freedom of choosing the conformal factor $f$. Below, we will explore this freedom and justify our non-affine $c$-function.
Using \ref{conditions} we can rewrite the expansions as follows:
\bea
\text{Affine}:&& \theta = (\text{constant} )\  \frac{ 2 b'}{b} \cr
\text{Non-Affine}:&& \tilde \theta = - 2 b', \ \text{with} \ f=\frac{b}{a}  \, .
\eea
Using the above expressions for expansion we can immediately find the c-function\footnote{For the affine choice we are using the proposal of \cite{Sahakian:1999bd} with an extra additive constant $c_2$ as in our proposal.}
\bea
\text{Affine}:&& c= \frac{c_0}{ 8 b'^3}+c_2 \cr
\text{Non-Affine}:&& c= \frac{c_1}{8 b'^3}+c_2 \, ,
\eea
Therefore, for this particular choice of $f$ both the affine and non-affine choices give the same $c$-function. As we have mentioned in the introduction this non-affinely defined $c$-function is arbitrary for different choice of the $f$. For instance for a different choice, $f=1$  we get 
\be
\tilde \theta = -\frac{ 2 a b'}{b}, \ \ \text{which gives} \ \ c= \frac{c_1 b^3}{8 a^3 b'^3}+c_2 \,.
\ee
At this point we will focus on our proposal of the non-affinely defined $c$-function. Note that in the above expression the harmonic function and the warp factor appear as a ratio with the same power. This will give us the correct boundary behaviour, which also implies that the function $f$ can have the form $\{\left (\frac{b}{a}\right)^n +\text{constant} \}$, where $n$ is any number including zero. This form of $f$ is model dependent. For a different example it will be different. 

The averaging of expansion ensures that we do not get any extra factor of $a$ or $b$ or AdS length scale in the c-function. For instance without the averaging we will get an extra factor of $\frac{1}{a b^2}$ for the $f=1$ case, which would make it diverge near the horizon. Additionally, it will generate a wrong power for the AdS scale. The cubic power of the derivative of $b$ makes sure that we get the correct power of the AdS radius. And the presence of the ratio of $(b/a)^3$ with the $b'^3$ is crucial for the finiteness and criticality of the c-function on the AdS. 

As discussed in the previous section the boundary is $AdS_5$ where the warp factor and the harmonic function becomes  
$b\rightarrow\frac{r}{L_{UV}}, \ a \rightarrow\frac{r}{L_{UV}}$
respectively. Therefore, the boundary (UV) value of the c-function simplifies to 
\be
c_{UV}= \frac{c_1 L_{UV}^3}{8}+c_2 \, .
\ee
On the other side of the RG flow, near the horizon, the harmonic function and the warp factor simplifies to  
%\be
$b\rightarrow \frac{r_h}{L_{IR}}, a \rightarrow \frac{r_h}{L_{IR}}$, where $r_h$ is the value of the radial coordinate near the horizon. 
%\ee
Consequently the IR value of the c-function becomes,
\be
c_{IR} = \frac{c_1 L_{IR}^3}{8}+c_2 \, .
\ee
From the CFT living on the boundary, we know that the UV and IR central charges are given as \cite{Myers:2010tj, brown1986, FORGE1989295}, 
\be
c_{UV}=\frac{\pi^2 L_{UV}^3} {G_5},\  c_{IR}=\frac{3 L_{IR}} {2 G_3},
\ee
where $G_5$ and $G_3$ are the five and three dimensioanl gravitational couplings respectively. Using these end point values of the c-function, we can solve for the constants $c_1 $ and $c_2$ in the definition of the c-function and get
\bea
c_1&=&\frac{8}{L_{UV}^3-L_{IR}^3} \left(   \frac{\pi^2 L^3_{UV} }{G_5} - \frac{3 L_{IR}}{2 G_3} \right) \cr
c_2&=&-\frac{L_{UV}^3}{L_{UV}^3-L_{IR}^3} \left(  \frac{\pi^2 L_{IR}^3 }{G_5} - \frac{3  L_{IR}}{2 G_3} \right).
\eea
Note that near the horizon, the 2d space that is in direct product with the $AdS_3$ might not be compact. But it can be made compact by quotienting out the phase space by the volume of the two dimensioanl space \cite{me}. If we consider the central charge as a measure of the degrees of freedom, then in principle on the IR end point we need to add the degrees of freedom corresponding to the 2d space with the $\text{CFT}_2$ central charge \cite{Behrndt:1998gr}. 
In the above computation we did not include this contribution. Including such term will not affect our proposal in any way. It would merely mean that we will get different values for the constants $c_1$ and $c_2$. 
%%%
%%%
\subsubsection*{$AdS_2$ Near Horizon Geometry}
Let's consider the case where the UV boundary is $AdS_4$ and the near horizon geometry is $AdS_2 \times \Sigma$. The computation is very similar as the $AdS_3$ case and we get the same expression for the expansion $\theta$. To get $AdS_2$ near the horizon we need to set $w(r)=b(r)$ and just as in the previous case we will consider a radial inwardly directed null ray. These rays are described by the null tangent vector
\be
k^\alpha = \left( a, -a^{-1},0,0\right) \, .
\ee
Then we find
\be
%\theta=
\nabla_{\mu} k^{\mu} =-(2 \frac{a}{b} b' + 2a' ) \,.
\ee
we also find that
\be
k^{\mu} \nabla_{\mu} k^{\nu} = -2 a' k^{\nu} \,.
\ee
Therefore 
\be
\tilde \theta =-2 \frac{a}{b} b' \,.
\ee
This is the same expression we obtained for the $AdS_3$ case with $f=1$ \footnote{If we choose, $f=\frac{b}{a}$, as in the $AdS_3$ case, we will get $\theta \sim b'$.}.
The c-function for this case is given as
\be
c= \frac{c_1 b^2}{4 a^2 b'^2}+c_2 \, .
\ee
On the $AdS$ boundary the above expression simplifies to
\be
c= \frac{c_1 L^2}{4}+c_2 \, .
\ee
The UV and IR central charges are not well defined. Therefore we can not evaluate explicitly the constants $c_1$ and $c_2$. 
In this case the boundary theory is a three dimensional CFT. For an odd dimensional CFT the central charges vanishes. However, for these theories the free energy of the CFT conformally mapped to a sphere is proposed by \cite{Klebanov:2011gs} to be monotonically decreasing function that is stationary at the UV and IR fixed points. More specifically, the conjectured c-function can be written as
\be
F = (-1)^{\frac{d􀀀-1}{2}} \log |Z| \, ,
\ee
where $|Z|$ is the $S^d$ partition function. For the solutions that interpolate between $AdS_4$ and $AdS_2$, this free energy provides the right boundary value data to determine the constants $c_1$ and $c_2$. 
%%%%

Recently Zaffaroni et al. \cite{Benini:2016rke} did an exact counting of microstates for the $AdS_4$ BPS black holes. It would be interesting to explore how our proposal can be incorporated with this development. 
%%%
\subsection*{Domain Wall Solutions}
Now, we will consider domain wall solutions in gauged supergravity theories that interpolate between two same dimensioanl $AdS$ spacetimes. The generic metric looks like
\be
ds^2=e^{2 A(r)} \left (-dt^2+\sum_{i}^{d-1}dx_i^2 \right)+dr^2 %=a^2 (-dt^2+dx_i^2)+dr^2 
\ee
To compute the expansion we will consider the following null vector
$k^{\mu}=\left(e^{-A}, -1, 0, 0,...\right)$. Then we get %expansion is then
\be
%\theta=
\nabla_{\mu} k^{\mu} =-4 A' \,.
\ee
Once again this random choice of null vector is not affine and a straightforward computation gives us the following:
\be
k^{\mu} \nabla_{\mu} k^{\nu} = -A' k^{\nu} \,.
\ee
Therefore we get the following affine expansion:
\be 
\tilde\theta %=2 \frac{a'}{a} 
=-3 A'(r).
\ee 
For an explicit illustration we will consider the five dimensional domain wall solutions. The c-function for these domain wall solutions becomes:
\be \label{domain}
c=\frac{c_1}{27 \ A'(r)^3}+c_2 \,.
\ee
At a fixed point where the geometry is $AdS$, the conformal factor is simply $ A(r) \rightarrow \frac{r}{L}$. This leads to the following values for the c-function:
\be \label{boundary}
c_{UV}= \frac{c_1 L_{UV}^3}{27}+c_2 \ \ \text{and} \  \ c_{IR}= \frac{c_1 L_{IR}^3}{27}+c_2 \, .
\ee
Using \ref{boundary} in \ref{domain}, we can immediately solve for the constants
\be
c_1= \frac{27 \pi^2} {G_5} \ , \ c_2=0 \, .
\ee
%%%%%%%%%%%%%%%%%%%%%%%%%%%%%%
%%%%%%%%%%%%%%%%%%%%%%%%%%%%%%
\section{$c$-Function and Holographic Entanglement Entropy}
In this section we will investigate how our proposal fits with the $c$-function proposed from the holographic entanglement entropy \cite {Casini:2006es, Myers:2012ed, Liu:2012eea}. Our approach would be more schematic than rigorous. Note that the universal term in the entanglement entropy plays the role of the central charge. We will use this information to justify that our geometric proposal is consistent with the entropic $c$-function. 

For convenience we will be using the Poincare coordinates and we will use the results of \cite{Hung:2011ta}. Consider a bulk spacetime that represents a RG flow for the boundary field theory. The bulk Einstein-scalar action is given by
\be
I \sim \int d^{d+1}x \sqrt{-g}  \left ( R - \frac{1}{2} G_{ab} \partial \phi_a \partial \phi_b -V  (\phi_a) \right) \, .
\ee
The RG flow can be induced by turning on the source term for some relevant operators or by developing a vacuum expectation value for certain scalar operators. The asymptotically AdS metric can be written as
\be
ds^2=\frac{L^2}{z^2} \left( f(z) (-dt^2 + dx_i^2) +dz^2\right) \, .
\ee
Here $z\rightarrow 0$ corresponds to the UV boundary. First we will find out the $c$-function based on our proposal. We will consider the following non-affine null vector:
\be
k^{\mu} =( z f , \ -z f, \ 0, ... ) \, .
\ee
Following the same steps as in the previous section we get the following $c$-function:
\be \label{cfun}
c= \frac{c_1 }{\left( 2-\frac{d}{2} \right) z f' + (d-1) f }+c_2 \, .
\ee
{\bf Holographic Entanglement Entropy:} Now we will explore the $c$-function from the entanglement entropy. We will employ the holographic prescription for computing the entanglement entropy \cite{Ryu:2006bv} and will be using the results obtained in\cite{Hung:2011ta}. The authors of \cite{Hung:2011ta} considered two flat parallel planes as the entangling surface. In the presence of a relevant deformation, the bulk surface is described with the profile $z = z(x_1)$ with the boundary conditions $z(x_1 = 0) = 0 = z(x_1 = l)$. The area of the entangling surface
\be
A = R^{d-2} L^{d-1} \int_0^l dx_1 \frac{ f ^{d/2-1}}{z^{ d-1}} \sqrt{ f + z'^2} \, .
\ee
By using this expression as an action for $z(x_1)$ we can determine the profile that will extremize the area. Since this action does not depend on $x_1$ explicitly, the conjugate momentum is conserved. %This conserved energy functional then yields the following equation:
%%%%
After minimizing the area we have
\be
S^{EE} \sim  \int_{\delta}^{z_*} \frac{ f^{d/2-1} }{z^{d-1}}  \left ( 1- \left(\frac{  f_* z^2 } { f z_*^2   } \right) ^{d-1}  \right ) dz  \, .
\ee
Here $z = \delta$ is the UV regulator and we will use the limit  $\delta \rightarrow 0$ in extracting the universal term. \\ \\ % Therefore we expand the integrand in powers of $z $ and evaluate only the term with 1/z. 
{\bf Comparison:}
For the purpose of illustration, we will consider $d=2$. The geometric $c$-function for this case simplifies to
\be \label{3d}
c= \frac{c_1}{z f'+f }+c_2
\ee
and the holographic entanglement entropy reads \cite{Hung:2011ta}
\be \label{entropy}
S^{EE}_3 \sim  \int_{\delta}^{z_*}  \left ( \frac{ 1 }{ z } - \left(\frac{  f_* } {  z_*^2   } \right)\frac{z}{f} \right ) dz \, ,
\ee
where $z_*$ is the maximum value of $z$ and $f_*=f(z_*)$. \\ \\
{\bf Near UV Behaviour:} For AdS boundary, $f \rightarrow 1$. Therefore, we can expand $f(z)$ near the UV fixed point ($\delta \rightarrow 0$) as
\be
f(z)= 1+\mu z^2+ ... \, ,
\ee
where $\mu$ is some mass scale. Using this expansion we get a constant $(c_1+c_2)$ for the $c$-function from our proposal
\ref{3d}. For the holographic entanglement entropy, the universal term (log-term) originates from the first term of the above integral. The coefficient of this universal term is a constant and it is the central charge of the dual CFT. \\ \\
{\bf Near  IR Behaviour:} For $AdS_3$, near the IR fixed point $f \rightarrow \frac{L}{L_{IR}}=\alpha >1$. Let's expand $f(z)$ near the IR fixed point as
\be
f(z)= \alpha- \bar \mu z^{-2}+ ... \, ,
\ee
where $\bar \mu$ is some mass scale. After plugging this expansion in \ref{3d}, we get a constant $\left (c_2+\frac{c_1 }{\alpha}\right)$ for the $c$-function. On the other side, from the holographic entanglement entropy \ref{entropy}, the coefficient of the universal term turns out to be  $ \left (C-\frac{C \bar\mu}{\alpha}\right )$, where $C$ is a constant. 

These similarities in structure for the near UV and IR behaviour indicate that our geometric proposal is consistent with the entropic proposal near the end points of the RG flow.
In fact, for the domain wall solutions these two proposals exactly match \cite{Casini:2006es, Myers:2012ed}. Note that, this section does not establish the relationship between the two proposals explicitly. However, it hints towards an underlying structure that is present in both proposals. An explicit realization of our $c$-function in the context of holographic entanglement entropy is left for future work.

%%%%%%%%%%%%%%%%%%%%%%%%%%%%%
%%%%%%%%%%%%%%%%%%%%%%%%%%%%%%
\section{Discussion}
The search for a $c$ functions addresses the old question about the irreversibility of the RG flow in a unitary quantum field theory. In the holographic context it explores if two conformal field theories defined on the boundary of the AdS space times are related by irreversible RG evolution. 

In this paper we try to address this issue and propose a holographic c function by using the non-increasing nature of the Raychaudhuri equation. Our key ingredient is the expansion parameter for the null congruences. Our proposal is boundary diffeomorphism invariant. Moreover, it has all the required properties such as the correct powers of the scale and the Newton constants. It is also stationary at the AdS boundaries which corresponds to the central charges for the CFT living there.  It is a generic proposal, will work in any space time dimension.  The only information we need is the metric of the space time. Our proposal also captures the scheme dependence of  the field theory $c$-function. %We also explored its feasibility for some examples. This finding would help us understanding the holographic RG better. 
The universal nature of our definition has a wide range of easy applicability to a variety of backgrounds as we have demonstrated in this paper. Further, this proposal might help us classify the information loss in RG flows. %in terms of how the future casual diamond's for a locus on a radial slice intersects null infinity in inverse proportion to the radial distance till at the black hole horizon, the asymptotic points are inaccessible to the radial slice. 

The universal nature of our definition offers a wide range of easy applicability for various backgrounds. In this paper, we have shown that this proposal can be applied in different scenarios, such as flow between two different dimensional CFTs, domain walls etc. It can incorporate these different scenarios in a single framework. Moreover, the simple structure of the proposal makes it generic and adaptable to any spacetime dimension. In recent years, this topic also gets a lot of attention in the context of holographic entanglement entropy, since the universal term of the entanglement entropy is identified as the central charge of the gauge theory. Besides, \cite {Casini:2006es}  has a proposal for the $c$-function based on the entanglement entropy which was later generalized for any dimensions by \cite{Myers:2012ed}. In this paper we showed that our geometric proposal is consistent with this entropic proposal. However, we do not have an explicit relationship between the two proposals yet. It would be interesting to see if our proposal can be realized by the holographic entanglement entropy. 
%Further, we hope this proposal might help us classify the information loss in RG flows. 

%%%%%%%%%%%%%%%%%%%%%%%%%%%%%%
%%%%%%%%%%%%%%%%%%%%%%%%%%%%%%
\section*{Acknowledgements} 
I am grateful to A. Osorio, S. Nampuri, B. Underwood, R. Carballo, T. Ali, R. Rahman and V. Jejjala for discussions and comments on the draft. I would also like to thank S. Khandakar for checking the draft. This research is supported by the Claude Leon Foundation. 

%%%%%%%%%%%%%%%%%%%%%%%%%%%%%%
%%%%%%%%%%%%%%%%%%%%%%%%%%%%%%
\bibliography{references}

\bibliographystyle{utphysmodb}

\end{document}